\documentstyle[12pt,epsfig]{article}
\def\Journal#1#2#3#4{{#1} {\bf #2}, #3 (#4)}
\def\NPB{{\em Nucl. Phys.} B}
\def\PLB{{\em Phys. Lett.}  B}

\def\PRD{{\em Phys. Rev.} D}
\def\ZPC{{\em Z. Phys.} C}

\def\drbar{\overline{\rm DR}}

\begin{document}

\baselineskip18pt
\begin{flushright}
MADPH-96-956 \\
hep-ph/9608382 \\
August, 1996
\end{flushright}

\vspace{2cm}
\begin{center}
\Large \bf 
Definition of $\tan\beta$ beyond 
tree-level\footnote{Talk at 1996 Annual Divisional Meeting (DPF'96) 
of the Division of Particles and Fields of the American Physical Society, 
Minneapolis, MN, Aug 10-15, 1996}
\end{center}
\vspace{10mm}

\begin{center}
\large Youichi Yamada
\end{center}
\vspace{0mm}

\begin{center}
\begin{tabular}{l}
{\it Physics Department, University of Wisconsin, Madison, WI 53706, USA}
\end{tabular}
\end{center}

%%%%% Abstract
\vspace{10mm}
\begin{center} 
\large Abstract
\end{center}
\begin{center}
\begin{minipage}{13cm}
\baselineskip=20pt
\noindent
%%%%% Here comes the abstract
We study the relation between different renormalization schemes for the 
parameter $\tan\beta$ in the Minimal Supersymmetric Standard Model. 
The contributions of the third-generation quark-squark loops 
to the differences between $\tan\beta$'s in several schemes are 
discussed. Their numerical differences are typically 
within several \%. 
\end{minipage}
\end{center}
\newpage
%\vspace{10mm}
\baselineskip=20pt
\normalsize
\setcounter{footnote}{0}

Some extensions of the Standard Model, including the Minimal 
Supersymmetric (SUSY) Standard Model\cite{mssm} (MSSM), 
contain two Higgs boson doublets. After the spontaneous breaking of 
SU(2)$\times$U(1) gauge symmetry, they get vacuum expectation 
values $(\langle H_1\rangle, \langle H_2\rangle)=(v_1,v_2)/\sqrt{2}$. 
Their ratio $\tan\beta\equiv v_2/v_1$ is a very 
important parameter in such models. In the MSSM, for example, masses 
and interactions of the SUSY particles and Higgs bosons 
depend\cite{mssm} on $\tan\beta$, 
which is very crucial for theoretical predictions, experimental 
search for new particles and a consistency test of the theory. 

In calculating radiative corrections to these observables, we have to fix 
a renormalization condition for $\tan\beta$. There are, however, 
no obvious criterion for the renormalization because $\tan\beta$ is not 
directly related to physical observables. Indeed, several 
renormalization schemes for $\tan\beta$ have been proposed in previous 
studies of the radiative corrections for Higgs 
bosons\cite{ay,brig,hh,cpr,dabel,a0def,hpdef}. In this paper, we 
demonstrate the relations among these definitions of $\tan\beta$ 
in the MSSM, by showing differences between them by the one-loop 
contributions from the third generation quarks and squarks. 

The MSSM Higgs sector lagrangian\cite{mssm} has five independent 
parameters $(m_1^2, m_2^2,$ $m_3^2, g_2, g_Y)$, among which $m_{1-3}^2$ 
are usually replaced with $(m_A, v_1, v_2)$. Here $m_A$ is the mass of the 
CP-odd Higgs boson $A^0$. We therefore 
need five inputs to fix these parameters. The parameters 
$(g_2, g_Y, \bar{v}^2\equiv v_1^2+v_2^2)$ are fixed by the electroweak 
gauge sector, as in the Standard Model. 
$m_A$ is usually chosen to be the pole mass of $A^0$. 
The remaining parameter $\tan\beta\equiv v_2/v_1$ is, however, 
not directly related with physical 
observables. In order to fix it, we must either 
give a procedure to define the 
``universal'', process-independent $\tan\beta$, or 
choose a physical input which depends on $\tan\beta$. 

In the former, we can define $\tan\beta$ by renormalized $v_{1,2}$. 
The complexity in this procedure arises from 
that to obtain physical results, we have to cancel the linear 
terms in the Higgs potential generated by radiative correction. 
These linear terms from tadpole contributions can be absorbed by shifting 
$(v_{1,2},m_{1-3}^2)$ but this operation is not unique\cite{cpr}. 
For example, we can just shift $v_{1,2}$ to cancel all tadpole 
contributions and then apply the modified minimal subtraction. 
In spite of its simplicity, $\tan\beta$ in this definition runs very rapidly 
with the renormalization scale\cite{effpot} and practically useless. 
Instead, we can shift both $v_{1,2}$ and $m_{1-3}^2$ 
to fulfil $\Delta v_1/v_1=\Delta v_2/v_2$, which is so far the most popular 
way in studying radiative corrections in the MSSM. 
As for the remaining divergence corresponding to 
$(Z_{H_2}/Z_{H_1})^{1/2}$, we can either apply the modified minimal 
subtraction\cite{ay,brig,hh} (called ``$\drbar$ scheme'' here) or 
fix it by imposing conditions on Higgs-gauge two-point 
function\cite{cpr,dabel} (``CPR scheme''). 

In the latter, the on-shell definitions of $\tan\beta$ are given by 
one physical observable and its tree-level relation with $\tan\beta$. 
Here we discuss the definitions\cite{a0def,hpdef} by leptonic partial 
decay widths of $A^0$ and of the charged Higgs boson $H^\pm$; 
\begin{equation}
\tan\beta_{A^0}\equiv
\frac{\bar{v}}{m_l}\left( \frac{8\pi}{m_A}\Gamma_{A^0\rightarrow l\bar{l}} 
\right)^{\frac{1}{2}},
\;
\tan\beta_{H^+}\equiv
\frac{\bar{v}}{m_l}\left( \frac{8\pi}{m_{H^+}}\Gamma_{H^+\rightarrow l^+\nu} 
\right)^{\frac{1}{2}}. \label{eq1}
\end{equation}
As for the renormalization of $\bar{v}$, 
we consider two on-shell 
definitions;\footnote{The effective coupling\cite{e2bar} 
$\bar{e}^2(m_Z^2)$ is used here to absorb large QED correction.} 
$1/\bar{v}^2=\sqrt{2}G_F$ or 
$1/\bar{v}^2=\bar{e}^2(m_Z^2)/(4m_W^2(1-m_W^2/m_Z^2))$. 
Each definition in Eq.~\ref{eq1} has then two versions. 
The another on-shell definition of $\tan\beta$ by the sfermion mass 
splitting\cite{ay,brig} does 
not work\cite{ay,sfermion} for large $\tan\beta$ and 
not discussed here. 

Here we consider the contributions of quarks and squarks in the third 
generation to the differences between $\tan\beta$'s given above. 
The counterterm for $\tan\beta\equiv t_\beta$ in each scheme then takes the 
following form\cite{ay,brig,hh,cpr,a0def,hpdef,adv}; 
\begin{eqnarray}
%\delta(t_\beta)/t_\beta |_{\drbar} &=&
%-3(m_t^2/v_u^2-m_b^2/v_d^2)\Delta/(16\pi^2), \\
\delta(t_\beta)/t_\beta |_{\drbar} &=&
-3(16\pi^2)^{-1}(m_t^2/v_2^2-m_b^2/v_1^2)(2/(4-d)-\gamma_E+\ln 4\pi), 
% error corrected
\nonumber \\
\delta(t_\beta)/t_\beta |_{\rm CPR} &=& (2m_Z)^{-1}(t_\beta+t_\beta^{-1})
{\rm Re}\Pi_{ZA}(m_A^2), \nonumber \\ 
\delta(t_\beta)/t_\beta |_{A^0} &=& {\rm Re}
[m_Z^{-1}t_\beta^{-1}\Pi_{ZA}(m_A^2)
+\Pi'_A(m_A^2)/2]+\delta\bar{v}/\bar{v}, \nonumber \\ 
\delta(t_\beta)/t_\beta |_{H^+} &=& {\rm Re}[-m_W^{-1}t_\beta^{-1}
\Pi_{W^-H^+}(m_{H^+}^2)
+\Pi'_{H^{\pm}}(m_{H^+}^2)/2]+\delta\bar{v}/\bar{v}. \label{eq2} 
\end{eqnarray}
$\Pi_{\cdots}(q^2)$ are unrenormalized two-point functions of 
Higgs and gauge bosons. 

Figure 1 shows numerical values of various $\tan\beta$ given in 
Eq.~\ref{eq2} as functions of $\tan\beta_{\drbar}(m_Z)$. 
As we included only the $(t,b,\tilde{t},\tilde{b})$ contributions, 
the relative differences between $\tan\beta$'s are large for small 
$\tan\beta$ (large top-Higgs Yukawa coupling). We can see that 
the differences are within several \%, the same order 
as the running of 
$\tan\beta_{\drbar}$ from $m_Z$ to $m_A$. 
Main parts of the differences come from the quark loops. 
As for the definitions by the leptonic decay widths, differences by the 
definition of $\bar{v}$ are quite significant. 

We have discussed the relation between $\tan\beta$'s in 
different renormalization schemes in the MSSM. 
We have shown that their numerical differences 
by the $(t,b,\tilde{t},\tilde{b})$ loops are typically within several \% 
for commonly used schemes. 
More complete study will be presented elsewhere\cite{adv}. 

%\section*{Acknowledgments} 
This research was supported in part by the U.S.~Department of 
Energy under Grant No.~DE-FG02-95ER40896 and in part by the 
University of Wisconsin Research Committee with funds 
granted by the Wisconsin Alumni Research Foundation.

\begin{figure}
\begin{tabular}{c}
\epsfig{figure=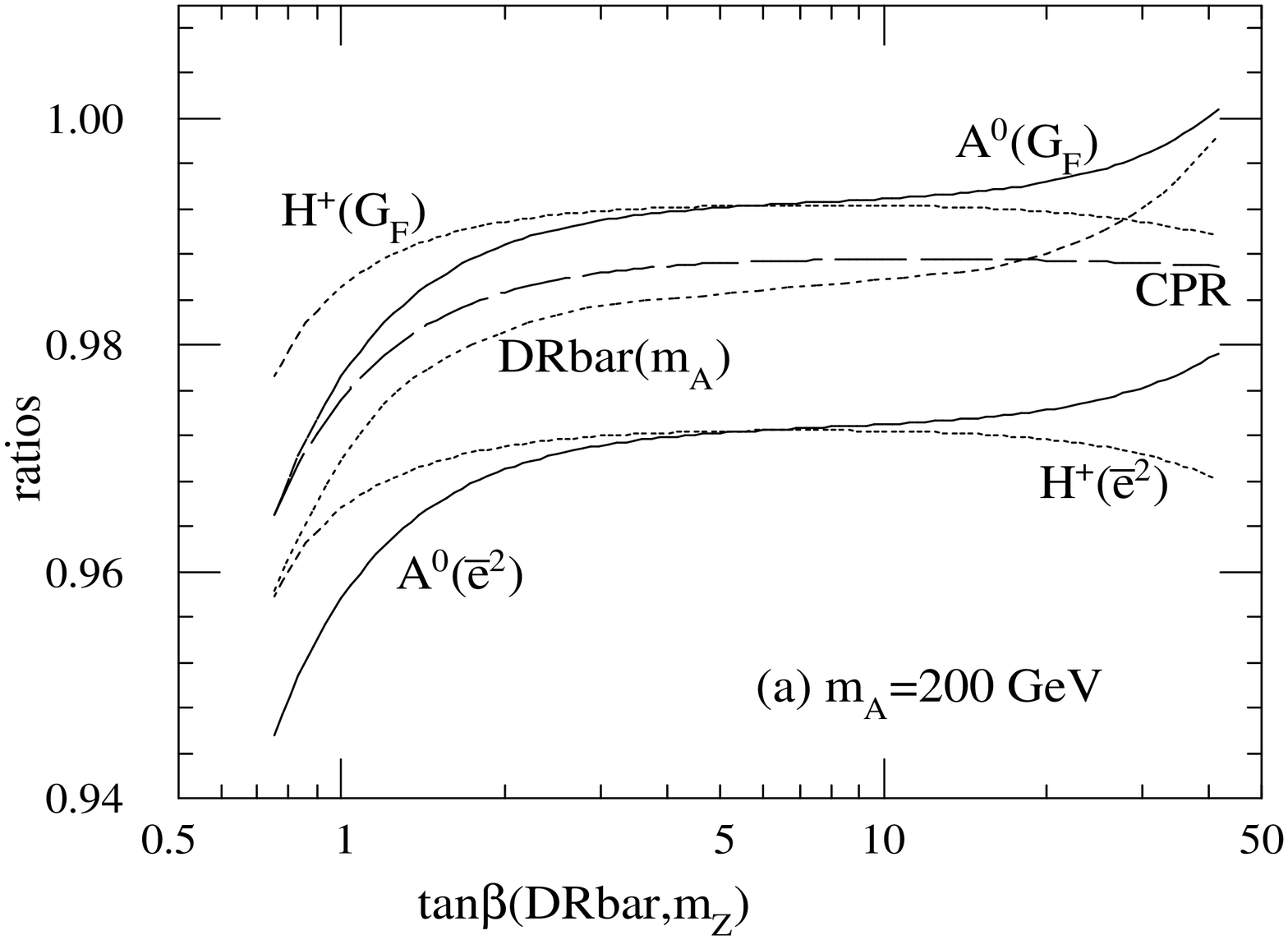,height=3.7in,angle=-90} \\
\vspace{10mm} \\
\epsfig{figure=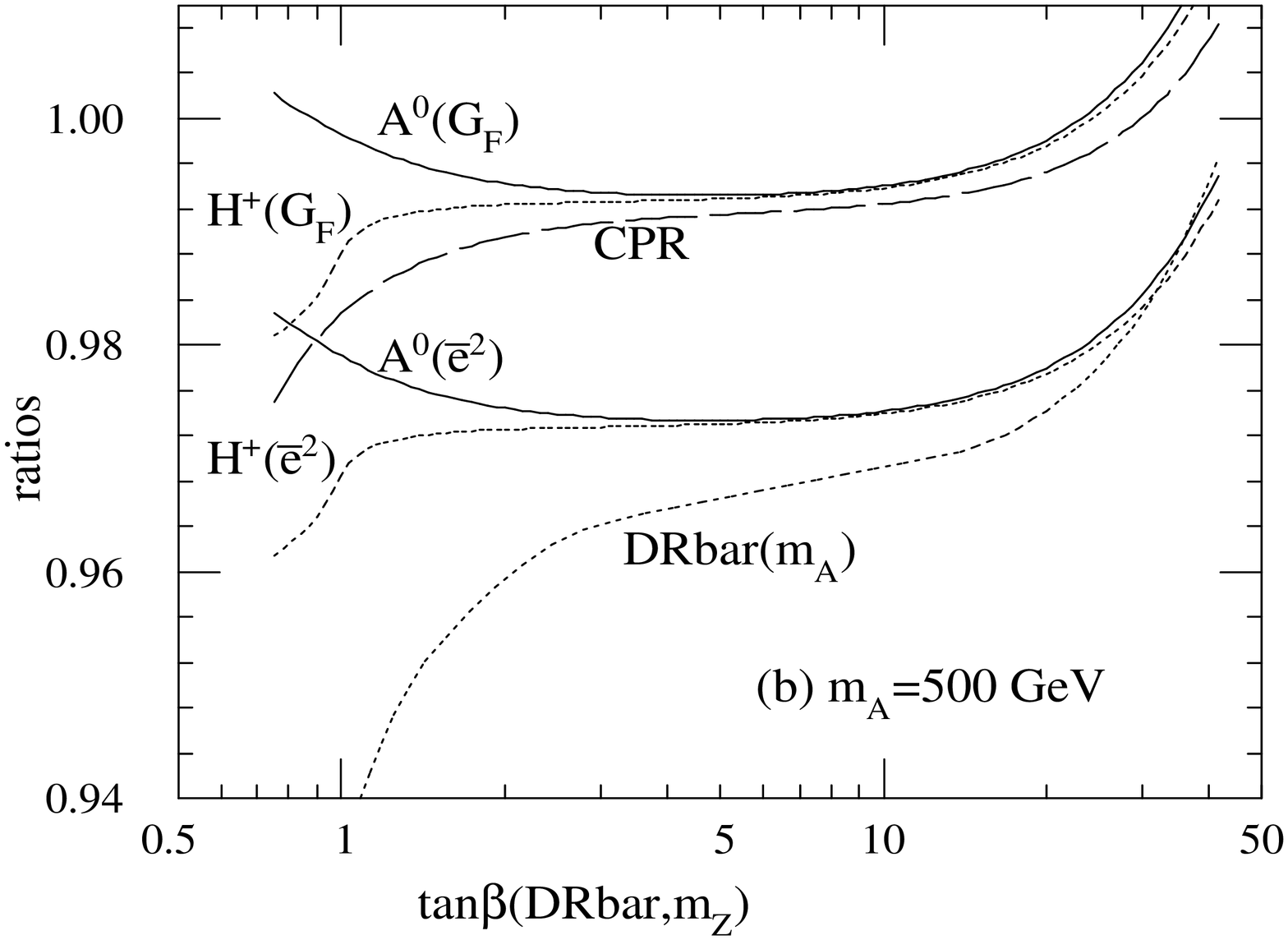,height=3.7in,angle=-90}
\end{tabular} 
\caption{Ratios of $\tan\beta$ in several definitions to 
$\tan\beta_{\drbar}(m_Z)$ for 
($M_{(\tilde{Q},\tilde{U},\tilde{D})}$, $\mu$, $A_{(t,b)}$)(GeV)= 
(300, 400, 0) and $m_A({\rm GeV})=$200 (a), 500 (b). \label{fig1}}
\end{figure}

\end{document}